\documentclass{article}
\usepackage{spconf,amsmath,graphicx}
\usepackage{times}
\usepackage{epsfig}
\usepackage{amsmath}
\usepackage{amssymb}
\usepackage{algorithm}
\usepackage{algorithmic}
\usepackage{enumerate}
\usepackage{multirow}
\usepackage{multicol}
\usepackage{arydshln}
\usepackage{color}
\usepackage{amssymb}

\usepackage{subfig}
\usepackage{subcaption}
\usepackage{caption}
\usepackage{float}
\usepackage{babel,blindtext}
\def\0{{\mathbf 0}}
\def\1{{\mathbf 1}}

\def\c{{\mathbf c}}

\def\e{{\mathbf e}}

\def\u{{\mathbf u}}
\def\v{{\mathbf v}}

\def\x{{\mathbf x}}

\def\A{{\mathbf A}}
\def\B{{\mathbf B}}
\def\C{{\mathbf C}}

\def\E{{\mathbf E}}

\def\I{{\mathbf I}}

\def\L{{\mathbf L}}

\def\P{{\mathbf P}}

\def\V{{\mathbf V}}
\def\W{{\mathbf W}}
\def\X{{\mathbf X}}
\def\Y{{\mathbf Y}}

\def\ie{{\textit{i.e.}}}

\def\cE{{\mathcal E}}

\def\cG{{\mathcal G}}

\def\cO{{\mathcal O}}

\def\cS{{\mathcal S}}
\def\cT{{\mathcal T}}

\def\cV{{\mathcal V}}
\def\cW{{\mathcal W}}

\def\bLambda{{\boldsymbol \Lambda}}

\title{Sparse Graph Learning from Sparse Data via Fiedler Number Maximization}
\name{{Bahar Oveisgharan$^\dag$, Gene Cheung$^\dag$, Andrew Eckford$^\dag$}
\thanks{The work of G. Cheung was supported by the Natural Sciences and Engineering Research Council of Canada (NSERC) RGPIN-2025-06252.}}
\address{$^\dag$York University, Canada  }

\begin{document}
%
\maketitle
\begin{abstract}
We aim to learn a sparse and connected graph from sparse data, where the number of observations $K$ can be substantially smaller than the signal dimension $N$ for signals $\x \in \mathbb{R}^N$, and the underlying distribution is unknown.
In this severely ill-posed setting, we incorporate Fiedler number (the second eigenvalue of the graph Laplacian matrix that quantifies connectedness) as a robust regularization term in the sparse graph learning objective.
We first develop a greedy algorithm that iteratively selects one edge globally for weakening / removal to reduce the objective, leveraging  eigenvalue perturbation theorems that bound the adverse effect of an edge change to the Fiedler number.
Next, we design a parallel variant, based on the Cheeger's inequality, that recursively partitions an input graph into two sub-graphs using an approximate Cheeger cut to distributedly find an optimal edge.
Simulation experiments show that Fiedler number maximization robusifies sparse graph estimates, outperforming previous sparse graph learning algorithms.
\end{abstract}

\section{Introduction}
\label{sec:intro}

\textit{Graph signal processing} (GSP) \cite{ortega18ieee,cheung18} studies mathematical tools such as transforms and wavelets to analyze discrete signals residing on irregular data kernels described by finite graphs. 
A basic premise in GSP is the existence of an underlying graph structure that captures pairwise similarities between the target signal's components.
If the graph is not known \textit{a priori}, there exist graph learning methods that compute a suitable graph structure from data \cite{dong19,Xia2021}.

Notably, statistical methods such as \textit{graphical lasso} (GLASSO) \cite{man0fried8}, CLIME \cite{cai11_clime}, and their many variants 
\cite{egilmez17,Kumar2019,bagheri21} assume as input an empirical covariance matrix $\bar{\C} = \frac{1}{K-1} \X \X^\top$---where observation matrix $\X \in \mathbb{R}^{N \times K}$ contains $K$ observations $\{\x_k\}$ as columns---and estimate a sparse inverse covariance (precision) matrix as output, interpreted as the graph Laplacian.
However, when the observation number is substantially smaller than the signal dimension, \ie, $K \ll N$, and the underlying distribution is not Gaussian, GLASSO- and CLIME-based methods perform poorly. 
%
%
This \textit{sparse data} scenario is not unusual in practice: collection of biological data may be costly or difficult (such as in live-animal experiments \cite{parhizkar2024signal}), or such data might be poorly labeled (such as in omics applications \cite{chang2024graph}). 

In this paper, we study the problem of learning a sparse connected graph from sparse data (\ie, $K \ll N$) for an unknown distribution. 
Given the severely ill-posed setting, we introduce  \textbf{Fiedler number  maximization}\footnote{\cite{bagheri2023} examined the effects on the Fiedler number when sparsifying a graph for GCN, but did not consider Fiedler number maximization as a regularization term in an optimization objective.}  as a robust regularization term into the sparse graph learning objective.
Fiedler number \cite{Chung1997}---also known as \textit{algebraic connectivity}---is the second eigenvalue of the graph Laplacian matrix that quantifies graph connectedness.
To attack the posed non-convex optimization problem, starting from an initial fully-connected graph, we devise a baseline greedy algorithm that iteratively selects one edge at a time for weakening / removal to reduce the objective.
We leverage eigenvalue perturbation theorems \cite{ipsen09} that bound the adverse effect of an edge change to the Fiedler number in each greedy step. 

To reduce complexity, we next design a parallel variant.
First, we build an initial graph from a \textit{minimum spanning tree} (MST) \cite{Cormen2009} plus $B$ extra ``information-rich" edges.
Next, based on the Cheeger's inequality \cite{Chung1997}, we recursively partition the graph into two sub-graphs using an approximate Cheeger cut to find an optimal edge in parallel.
\textit{To our knowledge\footnote{\cite{Tam2020} employed the Fiedler number for regularization in a loss function during end-to-end training of neural nets, while we insert the Fiedler number into a formal non-convex optimization to solve directly.}, we are the first to incorporate Fiedler number maximization for robust regularization in the sparse graph learning objective, specifically to promote connectedness under sparse data and unknown distribution conditions.}

Simulation experiments show that for distributions such as Gaussian mixture models (GMM) and multivariate t-distribution models (MtM), our graph learning algorithm outperforms GLASSO, CLIME, and extensions designed to target GMM and MtM specifically, demonstrating the advantages of Fiedler number maximization.

\section{Preliminaries}
\label{sec:prelim}
\subsection{GSP Definitions}
\label{subsec:defn}

A positive graph $\cG(\cV,\cE,\W)$ is defined by a node set $\cV = \{1, \ldots, N\}$ and an edge set $\cE$, where $(i,j) \in \cE$ means nodes $i,j \in \cV$ are connected with positive weight $w_{i,j} = W_{i,j} \in \mathbb{R}_+$.
We assume edges are undirected, and thus \textit{adjacency matrix} $\W \in \mathbb{R}^{N \times N}$ is symmetric. 
The \textit{combinatorial graph Laplacian matrix} is defined as $\L \triangleq \mathrm{diag}(\W \1) - \W \in \mathbb{R}^{N \times N}$, where $\mathrm{diag}(\v)$ returns a diagonal matrix with $\v$ along its diagonal.
Real and symmetric $\L$ is \textit{positive semi-definite} (PSD) if $w_{i,j} \geq 0, \forall i,j$, \ie, $\x^\top \L \x \geq 0, \forall \x$ \cite{cheung18}.

\subsection{Fiedler Number}

The graph Laplacian's second smallest eigenvalue, $\lambda_2(\L)$, is the \textit{Fiedler number}. 
This quantity has useful properties related to graph connectivity: $0 \leq \lambda_2(\L) \leq N$, where $\lambda_2(\L) = 0$ iff the graph is disconnected, and $\lambda_2(\L) = N$ iff the graph is complete. 
$\lambda_2(\L)$ is monotonically increasing as new edges are added to the graph. 
$\lambda_2(\L) \leq \kappa(\cG) \leq \eta(\cG)$, where $\kappa(\cG)$ is the \textit{node connectivity} of graph $\cG$, and $\eta(\cG)$ is the \textit{edge connectivity} of $\cG$. 
Thus, maximizing $\lambda_2(\L)$ is a useful proxy to optimize graph ``connectedness''. 




\subsection{Eigenvalue Perturbation Theorems}
\label{subec:perturb_thm}

\cite{ipsen09} first restated a known eigenvalue bound between $i$-th eigenvalue $\lambda_i$ of Hermitian matrix $\A$ and $\tilde{\lambda}_i$ of perturbed Hermitian matrix $\tilde{\A} = \A + \P$ based on Weyl's Theorem:
\begin{align}
|\lambda_i - \tilde{\lambda}_i| \leq \|\P\|_2 .
\label{eq:ebound1}
\end{align}
The disadvantage of \eqref{eq:ebound1} is that $\|\P\|_2$ can be large (resulting in a loose bound) and is independent of $i$. 

Denote the \textit{eigen-gap} at eigenvalue $i$ by
\begin{align}
\text{Gap}_i \triangleq \min_{j \neq i} |\lambda_i - \lambda_j| .
\end{align}
Denote by $\v_i$ the eigenvector for eigenvalue $\lambda_i$.
Lemma 3.1 in \cite{ipsen09} provides a tighter bound than \eqref{eq:ebound1} if $\text{Gap}_i > 2 \|\P\|_2$:
\begin{align}
|\lambda_i - \tilde{\lambda}_i| \leq \| \P \v_i\|_2 .
\label{eq:ebound_3}
\end{align}
A following result in Theorem 3.2 in \cite{ipsen09} states that for $\text{Gap}_i > \|\P\|_2$, the bound becomes
\begin{align}
|\lambda_i - \tilde{\lambda}_i| \leq  \sqrt{2} \, \| \P \v_i\|_2 .
\label{eq:ebound_2}
\end{align}

\subsection{Graph Sparsification via Fiedler Number}

%
Leveraging \eqref{eq:ebound_3},  \cite{bagheri2023} iteratively sparsifies a dense graph by greedily removing one edge at a time with the smallest $\ell_2$-norm $\|w_{m,n}\E^{m,n} \v_2\|_2$, where $w_{m,n}\E^{m,n}$ is the perturbation matrix that removes edge $(m,n)$ from $\L$, $\E^{m,n} \triangleq (\e_m - \e_n)(\e_m - \e_n)^\top$, and $\e_i$ is the $i$-th canonical vector. 
Second eigen-pair $(\lambda_2, \v_2)$ of $\L$ can be computed in $\cO(\text{nnz}(\L))$ (number of non-zero entries) via \textit{Locally Optimal Block Preconditioned Conjugate Gradient Method} (LOBPCG) \cite{Knyazev2001}.



\section{Problem Formulation}
\label{sec:formulation}
\subsection{Optimization Formulation}

We formulate an optimization for a symmetric adjacency matrix $\W \in \mathbb{R}^{N \times N}$ corresponding to a positive graph $\cG$ as follows.
Denote by $\X = [\x_1; \x_2; \ldots; \x_K] \in \mathbb{R}^{N \times K}$ the \textit{observation matrix} composed of $K$ observations $\{\x_k\}_{k=1}^K$ as columns.
We assume $K < N$, and thus $\X \X^\top$ is not full-rank.
To optimize matrix variable $\W$, we write
\begin{align}
\min_{\W} & ~~ \text{tr}(\X^\top \L \X) - \log \det(\L + \alpha \I) - \gamma \lambda_2(\L) + \mu \|\W\|_{0,\text{off}}
\nonumber \\
\text{s.t.} 
&~~ \L = \mathrm{diag}(\W \1) - \W, 
\label{eq:obj} \\
& ~~ \W = \W^\top, ~~ W_{i,i} = 0, \forall i, ~~W_{i,j} \geq 0, \forall i,j
\nonumber 
 \end{align}
%
where $\alpha, \gamma, \mu \in \mathbb{R}_+$ are parameters, and $\|\W\|_{0,\text{off}}$ is the $\ell_0$-norm of $\W$'s off-diagonal terms.
Fidelity term $\text{tr}(\X^\top \L \X)$  computes the smoothness of observations $\{\x_k\}$ with respect to (w.r.t.) the graph specified by $\L$  \cite{Dong2016}.
The first two terms together are interpreted as the log likelihood for a multivariate Gaussian distribution \cite{man0fried8}. 
The first constraint ensures a well-defined combinatorial graph Laplacian $\L$ given adjacency matrix $\W$. 
The remaining constraints ensure $\W$ is symmetric with zero diagonal entries (no self-loops) and non-negative off-diagonal entries (positive edge weights).
Overall, the objective is to find a sparse positive graph $\cG$ specified by $\W$, consistent with observations $\X$, while maximizing $\L+\alpha \I$'s determinant and $\L$'s second eigenvalue.

\subsection{Greedy Optimization Strategy}

Optimization \eqref{eq:obj} is non-convex; $\ell_0$-norm $\|\W\|_{0,\text{off}}$ in the objective is combinatorial, and the Fiedler number is non-convex w.r.t. $\W$ in general.
Our strategy is to optimize \eqref{eq:obj} greedily, where each greedy step \textit{maximally} decreases the objective.
Ignoring the sparsity term $\|\W\|_{0,\text{off}}$ for now, define $h(\W)$ as the sum of the first three terms in the objective given $\W$. 
At iteration $t$, we weaken a chosen edge $(m,n)$ by $\epsilon > 0$, \ie, $w_{m,n} \leftarrow w_{m,n} - \epsilon$. 
This is equivalent to subtracting perturbation matrix $\epsilon\P^{m,n}$ from adjacency matrix $\W$, \ie, $\W \leftarrow \W - \epsilon \P^{m,n}$, where
\begin{align}
\P^{m,n} = \e_m \e_n^\top + \e_n \e_m^\top .
\end{align}
$\P^{m,n}$ only has ones at entries $(m,n)$ and $(n,m)$.

Denote by $\nabla_\epsilon^{(m,n)}h(\W^{(t)})$ the (negative) change in objective $h(\W^{(t)})$ after weakening edge $(m,n)$ by $\epsilon$.
We select the best edge $(m^*,n^*)$ at iteration $t$ as
\begin{align}
(m^*,n^*) = \arg \min_{(m,n) \in \cE} \nabla_\epsilon^{(m,n)}h(\W^{(t)}) .
\end{align}
We next analyze how $\nabla_\epsilon^{(m,n)}h(\W^{(t)})$ can be computed.

\subsubsection{Effect of Edge Weakening on Trace Term}

Given $\L = \mathrm{diag}(\W\1) - \W$, perturbation $-\epsilon \P^{m,n}$ in $\W$ corresponds to perturbation $\L \leftarrow \L - \epsilon \, \mathrm{diag}(\P^{m,n}\1) + \epsilon \P^{m,n} = \L - \epsilon \E^{m,n}$.
We rewrite the trace term as $\text{tr}(\L \X \X^\top) = \text{tr}(\L \Y)$, where $\Y \triangleq \X \X^\top$.
Since $\text{tr}(\A\B) = \sum_{i,j} A_{i,j} B_{i,j}$ for real symmetric matrices $\A, \B$, the change in  $\text{tr}(\L^{(t+1)}\Y)$ (differential) w.r.t. $\epsilon$ is
\begin{align}
\partial \, \text{tr}(\L^{(t+1)}\Y) &= \partial \, \text{tr}((\L^{(t)} - \epsilon \E^{m,n})\Y) 
\nonumber \\
&= 2 Y_{m,n} - Y_{m,m} - Y_{n,n} \triangleq Z_{m,n}
\end{align}
where $Z_{m,n} \leq 0$, since $2 Y_{m,n} \leq Y_{m,m} + Y_{n,n}$ by the \textit{Cauchy-Schwarz inequality}. 
Thus, weakening an edge $(m,n)$ by $\epsilon$ always decreases the trace term.

\subsubsection{Effect of Edge Weakening on Determinant Term}

Using the \textit{Sherman-Woodbury identity}, the determinant of a rank-1 update $\A - \u \u^\top$ of matrix $\A$ is
\begin{align}
\det(\A - \u \u^\top) = \det(\A) \cdot (1 - \u^\top \A^{-1} \u) .
\end{align}
Given that the argument of our log determinant term is $\L + \alpha \I$, we can write

\vspace{-0.1in}
\begin{small}
\begin{align}
\det(\L^{(t)} + \alpha \I - \epsilon \E^{m,n}) &= \det(\L^{(t)} + \alpha \I)  \cdot 
\nonumber \\
& \!\!\!\!\! \underbrace{(1 - \epsilon (\e_m - \e_n)^\top (\L^{(t)} + \alpha \I)^{-1} (\e_m - \e_n) )}_{\eta_{m,n}} 
\nonumber \\
\partial \log \det(\L^{(t+1)} + \alpha \I) &= -\log \eta_{m,n} .
\label{eq:eta}
\end{align}
\end{small}\noindent

Computing $\eta_{m,n}$ directly requires a matrix inversion with complexity $\cO(N^3)$---costly though it is computed once for all $(m, n)$. 
Instead, we compute a spectral bound (\textit{majorization} \cite{Sun2017_MM}) using the first $K$ eigen-pairs $(\lambda_k,\v_k)$’s of $\L^{(t)}$:

\vspace{-0.1in}
\begin{small}
\begin{align}
\V_K \bLambda_K \V_K^\top + \alpha \I &\preceq \L^{(t)} + \alpha \I 
\label{eq:majorizer} \\
\V_K (\bLambda_K + \alpha \I)^{-1} \V_K^\top + \frac{1}{\alpha} (\I - \V_K \V_K^\top) &\succeq (\L^{(t)} + \alpha \I)^{-1}
\nonumber 
\end{align}
\end{small}\noindent
where $\bLambda_K$ is a diagonal matrix with $K$ smallest eigenvalues $\lambda_k$'s
along its diagonal, and $\V_K$ contains the corresponding eigenvectors $\v_k$’s
as columns.

\subsubsection{Effect of Edge Weakening on Fiedler Term}

We leverage the eigenvalue perturbation theorems in Section \ref{subec:perturb_thm} to bound the effect of edge weakening on the Fiedler term.
Suppose first that $\text{Gap}_2 > 2 \|\epsilon \E^{m,n}\|_2 = 4 \epsilon$, given $\|\E^{m,n}\|_2 = 2$. 
From \eqref{eq:ebound_3}, perturbing $\L^{(t)}$ by $-\epsilon \E^{m,n}$ means decreasing $\lambda_2$ by at most
$\|\epsilon \E^{m,n} \v_2\|_2$.
Thus,

\vspace{-0.1in}
\begin{small}
\begin{align}
\partial \lambda_2(\L^{(t+1)}) &\leq \epsilon \left\| \left[  \begin{array}{c}
v_{2,m} - v_{2,n} \\
v_{2,n} - v_{2,m} 
\end{array} \right]
\right\|_2
= \sqrt{2} \epsilon |v_{2,m} - v_{2,n}| .
\end{align}
\end{small}\noindent
Combining analysis for cases when $\text{Gap}_2 > 2\epsilon$ and $\text{Gap}_2 \leq 2\epsilon$ yields $\rho_{m,n} = \partial \lambda_2(\L^{(t+1)})$:

\vspace{-0.1in}
\begin{small}
\begin{align}
\rho_{m,n} &= \left\{ \begin{array}{ll}
\sqrt{2} \epsilon |v_{2,m} - v_{2,n}| & \mbox{if}~~ \text{Gap}_2 > 4 \epsilon \\
2 \epsilon |v_{2,m} - v_{2,n}|    & \mbox{elseif}~~ \text{Gap}_2 > 2 \epsilon \\
2 \epsilon    & \mbox{o.w.}
\end{array} \right. .
\label{eq:rho}
\end{align}
\end{small}

\subsubsection{Greedy Optimization Procedure}

We can conclude that weakening edge $(m,n)$ by $\epsilon$ results in the following change in $h(\W^{(t)})$:

\vspace{-0.1in}
\begin{small}
\begin{align}
\nabla^{(m,n)}_\epsilon h(\W^{(t)}) &= \epsilon Z_{m,n} - \log \eta_{m,n} + \gamma \, \rho_{m,n} - \mu \text{I}(w_{m,n} < \epsilon)
\label{eq:nabla_h}
\end{align}
\end{small}\noindent
where $\text{I}(\c)$ is an indicator function that equals to $1$ if clause $\c$ is true and $0$ otherwise.

We now design a \textit{greedy} procedure for iteration $t$ given a pre-chosen small step size $\epsilon > 0$:

\vspace{-0.05in}
\begin{small}
\begin{enumerate}
\item Compute $\v_2$ and $\text{Gap}_2$ for $\W^{(t)}$ via LOBPCG \cite{Knyazev2001}. 
\item For each edge $(m,n)$, compute $\nabla^{(m,n)}_\epsilon h(\W^{(t)})$ in \eqref{eq:nabla_h}.
\item Find edge $(m^*,n^*)$ with  smallest $\nabla^{(m,n)}_\epsilon h(\W^{(t)})$.
If $\nabla^{(m^*,n^*)}_\epsilon h(\W^{(t)}) < 0$: i) weaken edge $(m^*,n^*)$ by $\epsilon$, ii) goto step 2.
\end{enumerate}
\end{small}

\subsection{Algorithm Complexity}

Computation of $\v_2$ and $\text{Gap}_2$ via LOBPCG is $\cO(N^2)$ if $\W^{(t)}$ is dense with $\cO(N^2)$ entries.
For each one of $\cO(N^2)$ existing edge $(m,n)$, computing $\nabla^{(m,n)}_\epsilon h(\W^{(t)})$ in \eqref{eq:nabla_h} is $\cO(1)$, assuming that the spectral upper bound \eqref{eq:majorizer} for $(\L^{(t)} + \alpha \I)^{-1}$ to evaluate $\eta_{m,n}$ is computed once outside the loop, also with complexity $\cO(N^2)$.
Thus, the complexity for each edge weakening / removal is $\cO(N^2)$.
The root of this high complexity is a dense graph.

\section{Recursive Implementation}
\label{sec:implement}

\subsection{Initializing Connectivity}

Instead of a fully-connected initial graph with $\cO(N^2)$ edges, to reduce complexity we initialize a sparse graph with $\cO(N)$ edges, then we further weaken / sparsify edges to minimize objective \eqref{eq:obj}.
Given $\text{tr}(\L\Y) = \sum_i D_{i,i} Y_{i,i} - \sum_{i \neq j} W_{i,j} Y_{i,j}$, one connectivity strategy is to select large off-diagonal terms $Y_{i,j}$.
Specifically, we first construct a \textit{minimum spanning tree} (MST) with $N-1$ edges to ensure a connected graph, then add additional $B$ edges.
The intialization algorithm is

\begin{small}
\begin{enumerate}
\item Find the largest off-diagonal term $Y_{i,j}$ in $\Y$. 
Assign $w_{i,j} \leftarrow 1$, and initialize nodes set $\cS = \{i,j\}$. 
\item Find the largest off-diagonal term $Y_{i,j}$ connecting $i \in \cS$ and $j \in \cV \setminus \cS$.
Assign $w_{i,j} \leftarrow 1$, and add node $j$ to $\cS$. 
\item Repeat step 2 until $|\cS| = N$. 
\item Find the largest unselected off-diagonal term $Y_{i,j}$. 
Assign $w_{i,j} \leftarrow 1$.
\item Repeat step 4 until $N-1+B$ edges are selected.
\end{enumerate}
\end{small}\noindent
Step 1 to 3 is a variant of the Prim's algorithm for MST \cite{Cormen2009}, which is more efficient than the Kruskal's algorithm when the candidate edges (\ie, entries in $\Y$) are dense.

\subsection{Recursive Algorithm}

Given an initialized sparse graph, we design a recursive algorithm to identify an edge $(m^*,n^*)$ fo weakening / removal by partitioning a graph into two.
Specifically, to exploit parallelism, we recursively partition a graph into two leverage the \textit{Cheeger's inequality} \cite{Chung1997}. 


Denote by $\phi(\cG)$ the \textit{Cheeger constant} \cite{Chung1997} of a finite graph $\cG$, defined as
\vspace{-0.1in}
\begin{align}
\phi(\cG) = \min_{\cS \subseteq \cV, 0 < |\cS| \leq \frac{|\cV|}{2}} \frac{|\delta \cS|}{|\cS|}
\label{eq:cheeger_cut}
\end{align}
where $\cS$ (Cheeger cut) is a node subset of $\cV$, and $\delta \cS$ is the set of edges connected nodes in $\cS$ to $\bar{\cS} \triangleq \cV \setminus \cS$.
$\phi(\cG)$ specifies a partitioning of nodes $\cV$ into subsets $\cS$ and $\bar{\cS}$.
The \textit{Cheeger's inequality} \cite{Chung1997} relates the combinatorial graph notion in $\phi(\cG)$ with the spectral Fiedler number $\lambda_2$ as follows:
\begin{align}
\frac{\lambda_2}{2} \leq \phi(\cG) \leq \sqrt{2 \lambda_2 d_{\max}}
\label{eq:cheeger_bound}
\end{align}
where $d_{\max}$ is the largest node degree.
Given \eqref{eq:cheeger_bound}, the Cheeger cut $\cS$ in \eqref{eq:cheeger_cut} essentially identifies critical edges $\delta \cS$ that directly influence $\lambda_2$.


Leveraging \eqref{eq:cheeger_cut}, we recursively divide a graph $\cG$ into two sub-graphs, $\cG^1$ and $\cG^2$, as follows:

\vspace{0.1in}
\begin{small}
$(m^*,n^*) = \textbf{Partition}(\cG)$:
\begin{enumerate}
\item Given input graph $\cG$, if $|\cV| \leq V_{\min}$, compute $\nabla^{(m,n)}_\epsilon h(\W^{(t)})$ \eqref{eq:nabla_h} for all edges $(m,n) \in \cE$ and return the best one.

\item If $|\cV| > V_{\min}$, compute Cheeger cut $\cS$ \eqref{eq:cheeger_cut} to partition $\cG$ into $\cG^1$ with nodes $\cS$ and $\cG^2$ with nodes $\cV \setminus \cS$. 
Recursively call $(m^1,n^1) = \text{Partition}(\cG^1)$ and $(m^2,n^2) = \text{Partition}(\cG^1)$.  

\item Compare $\nabla_\epsilon^{(m,n)} h(\W^{(t)})$ for $(m^1,n^1)$, $(m^2,n^2)$ and Cheeger cut edges $\delta \cS$, and return the best one.

\end{enumerate}
\end{small}


\subsection{Algorithm Complexity}

Assuming the approximate Cheeger cut---with complexity $\cO(N)$ for a sparse graph---provides a roughly balanced graph partition into $\beta$ and $1-\beta$, total computation $\cW(N)$ for the original graph with $N$ nodes and $\cO(N)$ edges is
\begin{align}
\cW(N) = \cW(\beta N) + \cW((1-\beta)N) + \cO(N) .
\end{align}
Based on the \textit{Master Theorem} \cite{Cormen2009}, the complexity is $\cW(N) = \cO(N \log N)$. 
If each recursive call is computed in parallel by separate processors (best-case scenario), then running time $\cT(N)$ is
\begin{align}
\cT(N) = \max \left( \cT(\beta N), \, \cT((1-\beta)N) \right) + \cO(N),
\end{align}
resulting in $\cT(N) = \cO(N)$.
In practice with finite number of processors, running time is between $\cO(N)$ and $\cO(N \log N)$ for each graph edge weakening / removal.

\section{Experiments}
\label{sec:results}

Following the synthetic-data protocol of \cite{bagheri21}, we generate a ground-truth sparse graph, compute its (loop-free) Laplacian $\L$, and define the precision matrix as $\boldsymbol{\Theta}=\L+\rho\I$ (equivalently, covariance $\C=(\L+\rho\I)^{-1}$) with $\rho=0.5$. 

To evaluate the robustness of our method under non-Gaussian conditions while preserving the graph-induced dependencies, we generate synthetic signals using two non-Gaussian sampling schemes based on a prescribed precision matrix $\boldsymbol{\Theta}\succ 0$ (with covariance $\boldsymbol{\Theta}^{-1}$): \textbf{(i) a Gaussian mixture model (GMM)} with shared covariance across components, which yields a non-Gaussian marginal distribution, and \textbf{(ii) a multivariate Student’s $t$ distribution}, which produces heavy-tailed non-Gaussian signals. Repeating either procedure for $K$ draws forms the signal matrix $\mathbf{X}\in\mathbb{R}^{N\times K}$.

\begin{figure}[t]
    \centering
    \includegraphics[width=0.95\columnwidth]{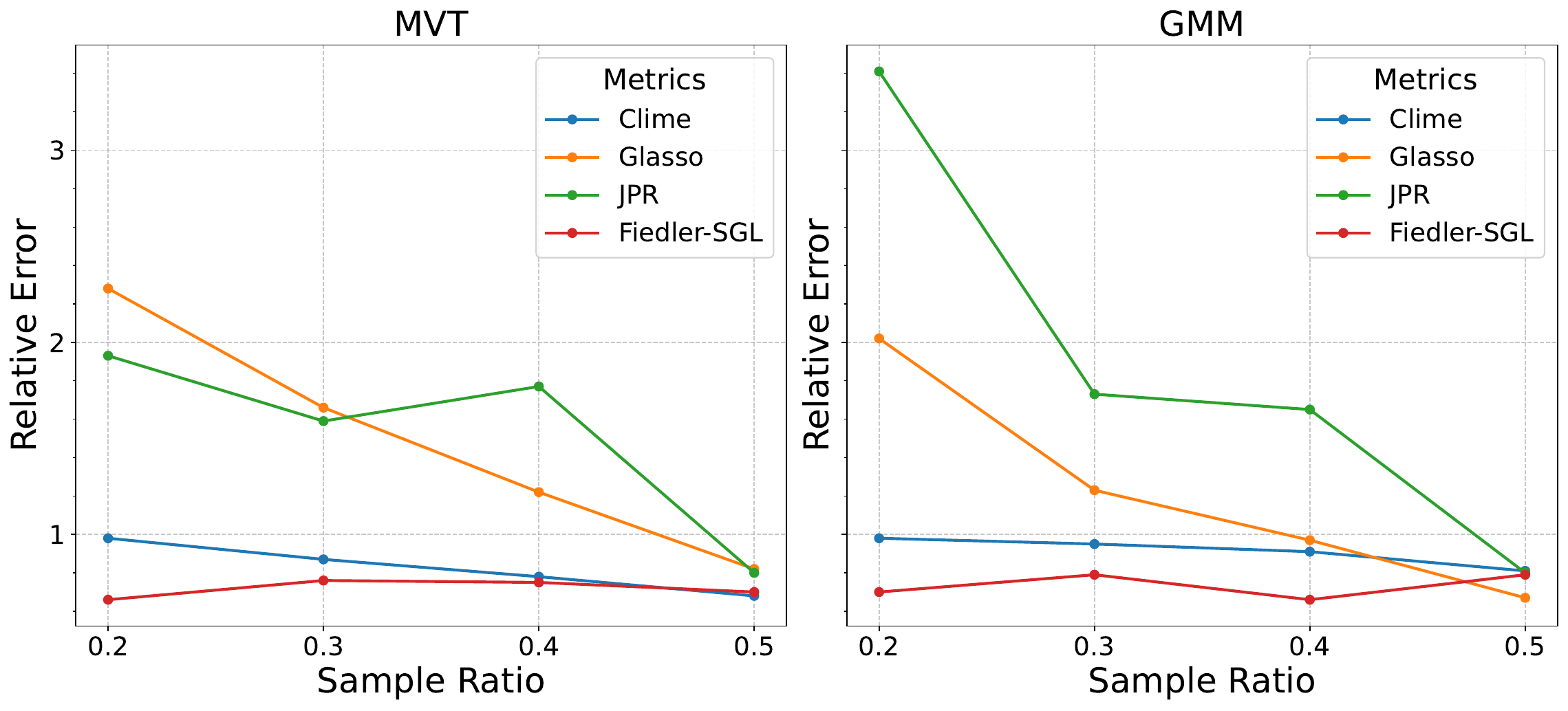}
    \vspace{-0.1in}
    \caption{Relative error versus sample ratio for two data-generation settings.}
    \label{fig:two_plots}
\end{figure}

We set the step size to $\epsilon=0.01$; smaller values typically require more iterations (increasing runtime) but can yield finer updates. We use $\gamma=0.5$ to control the strength of the connectivity regularization (larger $\gamma$ enforces stronger connectivity) and set $\mu=0.2$ to promote sparsity in the learned graph (larger $\mu$ yields sparser solutions).

Figure~\ref{fig:two_plots} compares our method, Fiedler-regularized Sparse Graph Learning  \textbf{(Fiedler-SGL)}, with GLASSO~\cite{man0fried8}, CLIME~\cite{cai11_clime}, and Joint Precision Regression (Graph-JPR)~\cite{lee2015jpr} as the number of samples $K$ varies on graphs with a fixed size $N=30$. Results are shown versus the sample ratio $K/N$ under two non-Gaussian data-generation settings (left: GMM, right: MVT). In the undersampled regime ($K \ll N$), Fiedler-SGL consistently achieves the lowest reconstruction error. We quantify accuracy using the relative error $\mathrm{RE}=\|\widehat{\mathbf{W}}-\mathbf{W}^\star\|_F/\|\mathbf{W}^\star\|_F$ (smaller is better).

Table\;\ref{tab:runtime} highlights a substantial runtime advantage of the proposed sparse initialization + recursion pipeline over the baseline greedy method. For $N=20$, the average runtime drops from $8.151$ s to $1.836$ s (approximately $4.4\times$ faster), and for $N=30$ it decreases from $42.511$ s to $5.816$ s (approximately $7.3\times$ faster), with the gap widening as $N$ grows. This improvement is driven by two design choices. First, {sparse initialization} constructs an initial connected graph with only $\mathcal{O}(N)$ edges, rather than starting from a dense $\mathcal{O}(N^2)$ candidate set, which significantly reduces both computation and memory. Second, the recursive variant structures the procedure into smaller sub-problems, allowing better reuse of intermediate computations and enabling parallel execution of independent loops, leading to more efficient utilization of available compute resources.

\begin{table}[t]
\caption{Average runtime in seconds.}
\label{tab:runtime}
\centering
\vspace{-0.1in}
\begin{small}
\begin{tabular}{lrr}
\hline
$N$ & Greedy (s) & Sparse init.\ + recursion (s) \\
\hline
20 & 8.151  & 1.836 \\
30 & 42.511 & 5.816 \\
\hline
\end{tabular}
\end{small}

\end{table}

\section{Conclusion}
\label{sec:conclude}

We address the severely ill-posed problem of learning a sparse graph from sparse data when the underlying distribution is unknown. 
To robustify the graph estimate, we maximize in addition the Fiedler number in the optimization objective via a greedy algorithm weakening / removing an edge at a time, where the effects of an edge weakening on the Fiedler number is bounded using eigenvalue perturbation theorems.
To exploit parallelism and reduce complexity, we design a recursive algorithm based on the Cheeger's inequality to partition a graph to identify an edge for weakening. 
Synthetic experiments show that at the sparse data regime when $K \ll N$, our graph learning method noticeably outperforms SOTA schemes for the student's t-distribution.

\begin{footnotesize}
\bibliographystyle{IEEEbib}
\bibliography{ref2,aweref}
\end{footnotesize}

\end{document}